\documentstyle[psbox]{WORKSHOP}
\begin{document}%

\title{
Hard X-Ray sources observed by INTEGRAL/IBIS and their science \\
}
\author{Pietro Ubertini on  behalf of the IBIS Survey Team  \\
\\[12pt]  
%
 IASF-Roma, INAF, Via Fosso del Cavaliere 100, 00133 - Rome, Italy \\
%
{\it E-mail: pietro.ubertini@iasf-roma.inaf.it}
}

\abst{ After more than 5.5 years of in flight lifetime, the ESA
space observatory INTEGRAL is depicting a new scenario in the soft
gamma Ray domain. With the observation and discovery of more than
400 hard X-Ray sources has changed our view of a moderately
crowded and dark sky basically populated by "peculiar" and erratic
sources. The new high energy sky is also composed by a large
variety of "normal" though very powerful objects, characterized by
new accretion and acceleration processes. Among the 421 sources
detected in the energy range 17-100 keV the 3$^{rd}$ INTEGRAL/IBIS
catalogue includes 41\% galactic accreting system, 29\%
extragalactic objects, 8\% of other types, and 26\% not classified
i.e. of unknown origin. If compared to to previous IBIS/ISGRI
surveys it is clear that there is a continuous increase of the
rate of discovery of HMXB, an increase in the number of AGN
discovered, including a variety of distant QSOs, basically due to
the increased exposure away from the Galaxy Plane, while the
percentage of sources without an identification has remained
constant. INTEGRAL, by the end of Y6, will complete the Core
Programme observation due to provide an almost constant exposure
of the whole Galaxy Plane and will be fully open to the scientific
community for Open Time and Key Programme observations from the
beginning of 2009. In this paper we shortly review the main
outcome of  the 3$^{rd}$ INTEGRAL/IBIS catalogue (cat3) including
an excursus of the INTEGRAL high energy sky with particular regard
to sources emitting at high energy, including Low and High Mass
X-Ray Binaries (LMXB \& HMXB),  Supergiant Fast X-Ray Transients
(SFXT), Pulsar Wind Nebulae (PWN) and Active Galactic Nuclei
(AGN).}

\kword{Subject headings: Gamma-rays, observations, surveys}

\maketitle
\thispagestyle{empty}

\section{INTRODUCTION}

The INTEGRAL ESA gamma-ray observatory has already carried out
more than 5 years of Core, Key and Guest Observer Programme
observations in the energy range from 5 keV - 10 MeV (Winkler et
al. 2003). The IBIS coded mask imager is optimised for scientific
observations over a wide field of view (${\sim}$1000 squared
degree) with excellent imaging and spectroscopy capability
(Ubertini et al. 2003), to provide survey type data, including
detailed images of the whole sky, time profile and spectra of all
the sources detected with a sensitivity better than 1 mCrab in the
central radian.

\section{The INTEGRAL Observatory}

The ESA INTEGRAL (International Gamma-Ray Astrophysics Laboratory)
observatory was selected in June 1993 as the next medium-size
scientific mission within ESA Horizon 2000 programme.
\begin{figure}
\centering
 \psbox[xsize=6cm]{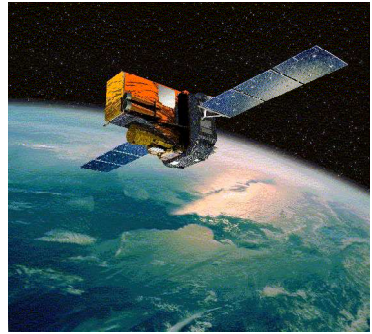}
 \vspace{-4cm}
\caption{The INTEGRAL Observatory (ESA Courtesy).}
 \label{fig:ESA}
\end{figure}
INTEGRAL (Winkler et al. 2003), shown in Figure \ref{fig:ESA}, is
dedicated to the fine spectroscopy (2.5 keV FWHM @ 1 MeV) and fine
imaging (angular resolution: 12 arcmin FWHM) of celestial
gamma-ray sources in the energy range 15 keV to 10 MeV. While the
imaging is carried out by the imager IBIS (Ubertini et al. 2003),
the fine spectroscopy if performed by the spectrometer SPI
(Vedrenne et al. 2003) and coaxial monitoring capability in the
X-ray (3-35 keV) and optical (V-band, 550 nm) energy ranges in
provided by the JEM X and OMC instruments (Lund et al. 2003;
Mass-Hesse et al. 2003). SPI, IBIS and Jem-X, the spectrometer,
imager and X-ray monitor are based on the use of coded aperture
mask technique that is a key feature to prove images at energy
above tens of keV, where photons focusing become impossible using
standard grazing technique. Moreover, coded mask feature the best
background subtraction capability because of the possibility to
observe at the same time the Source and the Off-Source sky. This
is achieved at the same time for all the sources present in the
telescope field of view. In fact, for any particular source
direction, the detector pixels are split into two sets: those
capable of viewing the source and those for which the flux is
blocked by opaque tungsten mask elements. This very well
established technique is discussed in detail by (Skinner \&
Connell 2004) and is extremely effective in controlling the
systematic error in all the telescope observation, working
remarkably well for weak extragalactic field as well for crowded
galactic regions, such as our Galaxy Center. The mission was
conceived since the beginning as an observatory led by ESA with
contributions from Russia (PROTON launcher) and NASA (Deep Space
Network ground station). INTEGRAL was launched from Baikonur on
October 17$^{th}$, 2002 and inserted into an highly eccentric
orbit (characterized by 9000 km perigee and 154000 km apogee). The
high perigee in order to provide long uninterrupted observations
with nearly constant background and away from the electron and
proton radiation belts. Scientific observations can then be
carried out while the satellite being above a nominal altitude of
60000 km, while entering the radiation belts, and above 40000 km,
while leaving them. This strategic choice ensure about 90\% of the
time is used for scientific observations with a data rate of
realtime,108 kbps science telemetry received from the ESA station
of Redu and NASA station of Goldstone. The data are received by
the INTEGRAL Mission Operation Centre (MOC) in Darmstadt (Germany)
and relayed to the Science Data Center (Courvoisier et al. 2003)
which provide the final consolidated data products to the observes
and later archived for public use. The proprietary data become
public one year after distribution to single observation PIs.
\begin{figure}[h]
 \centering
\psbox[xsize=5cm]{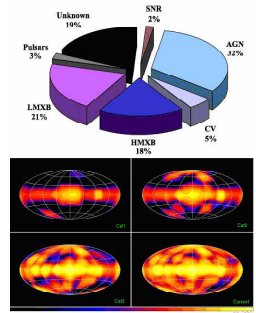}
\vspace{-1.5cm}
 \caption{Top:distribution of the high energy sources
of the 3$^{rd}$ IBIS Catalogue: for each source is also produced
the spectrum and the light curve. Bottom: IBIS Sky coverage
evolution vs time.}
 \label{fig:SKY}
\end{figure}

\begin{figure}[h]
 \centering
 \psbox[xsize=3.8cm]{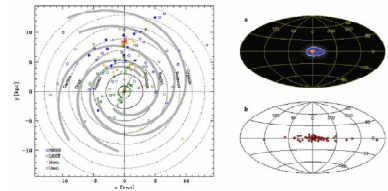}
 \vspace{-1.5cm}
 \caption{Left: distribution of IBIS sources in the Galaxy. Right-top:
discovery of asimmetric distribution of positron 511 keV
annihilation in the Galactic disk. Right-bottom: IBIS LMXB
distribution.}
\label{fig:SOURCE}
\end{figure}

\section{The INTEGRAL/IBIS sky}
\subsection{The Catalogue}
In order to give maximum access to the whole scientific
INTEGRAL/IBIS data set  a catalogue, comprising all the detected
high energy sources is generated roughly on an annual basis (Bird
et al, 2004, 2006, 2007). The last of them (the 3$^{rd}$, namely
THE 3$^{rd}$ IBIS/ISGRI SOFT GAMMA-RAY SURVEY CATALOG) has been
produced using all the IBIS data available up the and of May 2006
(40 Ms exposure time). It contains 421 sources, detected above
4.5${\sigma}$ in the energy range 18-100 keV; in Figure
\ref{fig:SKY} is shown the distribution of the sources (top panel)
and the INTEGRAL/IBIS Sky coverage evolution as per cat 1, 2 and
3; the bottom panel on the right is the coverage at the end of Y5.
As expected the majority of the IBIS sources are located at low
galactic latitudes, due to the basic fact that INTEGRAL is
frequently observing in the Galactic Plane and the sky coverage is
far for uniform, as shown in Figure \ref{fig:SKY}. The source
identification process is based on a multi wavelength approach,
taking into consideration the Radio, IR, X-ray and higher energy
archival data. A robust programme of optical and IR observation
campaigns has been activated just after the INTEGRAL launch and
had been very successful (Masetti et al. 2006a,b,c,d and 2008).
Among the sources that have been firmly identified are 171
associated with galactic binary accreting systems (41\%), 122  are
extragalactic sources (29\%), 15 belong to different classes of
high energy emitters. As many as 113 (26\%) are still not
identified. The sources belonging to our Galaxy are sub-divided
into 21 CV systems (9 of which are new detections with emission
extending up to 100keV), 65 high-mass X-ray binaries (HMXB) and 78
low-mass X-ray binaries (LMHB). IBIS continues to detect Low Mass
systems X-Ray Binaries even if  the rate of discovery is much
lower than for the high mass systems. In particular, the High Mass
Binary includes 19 new IBIS sources which have been identified
with Be systems on the basis of their spectral characteristic
and/or transient behavior. A distinct type of new objects,
increasing regularly in the INTEGRAL/IBIS sample are the Super
Giant fast X-ray transient (SFXT). Apart from the detailed and
unbiased study related to the IBIS/ISGRI catalogue production
several outstanding studies and results of wide astrophysical
interest have been carried out with INTEGRAL/IBIS. An overview,
even if superficial, is behind the scope of this short overview.
Just as an example in Figure \ref{fig:SOURCE} (left panel) is
shown the distribution of the IBIS sources in the Galaxy: circles
are HMXB, squares are LMXB, triangles are Miscellaneous and
inverted triangles are unclassified sources (Bodaghee et al.,
2007). Of complete different nature the discovery of asimmetric
distribution of positron annihilation in the Galactic disk based
on the SPI outstanding spectroscopic capability and IBIS high
angular resolution (Weidenspointner et al., 2008) as shown in
Figure \ref{fig:SOURCE} (left panel): the IBIS LMXB distribution;
(right panel) the SPI 511 keV emission. This result clearly
demonstrate the SPI-IBIS complementarity.

\begin{figure*}[t]
 \centering
\psbox[xsize=10cm]{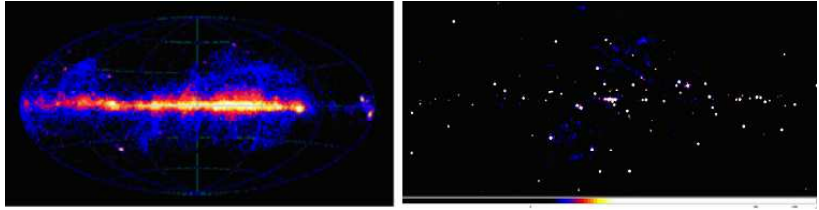}
\vspace{-8cm}
 \caption{The AGILE, E$>$100 MeV, High Energy Sky, dominated by the Disk diffuse emission
  (left) and the IBIS soft gamma Ray  sky, E$>$20 keV, dominated by discrete sources. }
  \label{fig:GAMMA}
\end{figure*}

\subsection{The IBIS sky: E$>$100 keV }
 The data accumulated with INTEGRAL/IBIS at energies
above 100 keV in the first 3.5 years of operative life, have been
discussed in Bazzano et al. (2006). This data set provided the
first IBIS sources  catalog characterised by very hard spectral
components, listing 49 sources detected with a significance above
4 in the 100-150 keV energy range of which only 14 seen above 150
keV. Black Holes and accreting Neutron Stars systems mainly
populate the high energy sky but extragalactic source are also
contributing with 20\% of the sample. In the higher energy range
($>$150 keV) the main emitting sources are Black Hole
candidates/microQSOs.  A larger sample is due to be finalized and
the sample consist now of about 100 sources. All classes are
included with the majory being galactic sources. The IBIS sample
is consistent with the SPI resolved point sources in the 100-200
keV band (Bouchet et al., 2008). Also, an indication of the
possible association between the the observed asymmetry in the 511
keV line emission from the galactic disk with X-ray Binary, mainly
LMXB, has been suggested by Weidenspointner et al., 2008 (cfr
paragraph 3.1). It was also noticed by the same author that
similar asymmetry is derived when taking into account the LMXB
distribution as from the 3$^{rd}$ IBIS catalog above 20 keV (Bird
at al., 2007) and even more evident when using the sources above
100 keV. Finally, in Figure \ref{fig:GAMMA} is shown the AGILE
High Energy Sky for E$>$100 MeV (Tavani et al. 2008), dominated by
the Disk diffuse emission (left) and the IBIS soft gamma Ray  sky,
E$>$20 keV, dominated by discrete sources (right).
\begin{figure}[h]
 \centering
 \psbox[xsize=6cm]{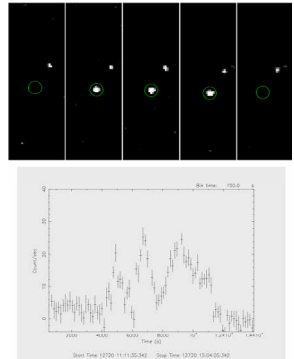}
 \vspace{-2cm}
 \caption{Short time variability of XTE J1739-302.}
\label{fig:SHORT}
\end{figure}
\subsection{The keV to TeV connection}
Another significant class of sources strongly emitting in the soft
gamma-ray range as well as at very high energies ($>$TeV) are the
Pulsar and Pulsar Wind Nebulae (PWN). While there is no room in
this short paper to properly address this  topic, it is worth to
mention that IBIS has detected more than 10 of those high energy
sources, most of them associated with HESS/MAGIC high energy gamma
ray emission (For a comprehensive review see Aharonian et al.
2005, Albert et al., 2007, Malizia et al. 2005, Ubertini et al.
2005, Dean et al. 2008, Ubertini et al., 2008 and references
therein).

\subsection{FXT sources and Supergiant HMXB}
One of the two new class of high energy emitters discovered by
INTEGRAL belong to the category of Fast Transient X-Ray sources
(FXT), the other one correlated with strongly absorbed binary NS
of which J16318-4848 is the archetype (Walter et al., 2006). Even
if one of the most common characteristics of the high energy sky
is the source variability (from ms to Ms), very little was known
before the INTEGRAL launch on the class of sources that were
identified as FXT. The  basic information available were
correlated with their location, off the Galaxy plane, and their
seldom large outbursts lasting less than a day. The only un-biased
survey of those class of objects was performed by BeppoSAX with
the Wide Field Camera detectors, taking advantage of their large
field of view (2$\times$40$\times$40 squared degree looking at 180
degree each other) and the total mission span along more than 6
years. BeppoSAX detected almost 50 different objects that
exhibited a bimodal distribution in their duration. The first type
is now known to be associated with the so-called X-Ray flashes,
connected with Gamma-ray bursts type event (D'Alessio et al.,
2006) and the second one, connected with active coronal star. More
recently, INTEGRAL has changed the BeppoSAX scenario, thanks to
the extended high energy response of the IBIS telescope, via the
discovery of the so-called Supergiant Fast X-Ray Transient (Sguera
et al., 2005, 2006, 2007, 2008; Neguerela et al., 2006). This new
class of high energy emitters, SFXTs, has been optically
identified with HMXB hosting an early type supergiant star. The
surprise was that this kind of classical Supergiant HMXB (called
SGXBs), discovered more than 40 years ago, were known to be
persistent X-ray emitters with an emission ranging from
$10^{36}-10^{38}erg$ $s^{-1}$. In fact, the SFXTs have a quiescent
luminosity from 1000 to 1000,000 times smaller and therefore
undetectable most of the time with current X-Ray monitoring
experiment. Conversely, they occasionally have fast X-ray activity
lasting from hours to a day (or more) with peak luminosity of the
order of 10$^{36}-10^{37} erg$ $s^{-1}$. INTEGRAL revealed flares
characterized by short duration peaks (ten of minutes) and steep
spectra above 20 keV, with photon power law index between 2-3.
Alternatively, the spectra can be fitted with thermal
bremsstrahlung models (kT comprised between 10 and 35 keV). In
Figure \ref{fig:SHORT} (top panel) is shown a "classical"
detection of the IBIS telescope: the images are related to the
detection of a fast X-ray outburst of XTE J1739$-$302 in the
energy range from 20 to 30 keV. Each panel has a duration of about
2000s and the source is visible only in the 3 central panel, for a
total flare duration of less that one hour. In the first and last
half hour observation is well below the flux at the peak and was
not detected at all. The stable serendipity source in the top part
of the image is the BHC 1E 1740.7-2942. In the right panel is
shown the IBIS/ISGRI light curve integrated over the energy range
from 20 to 60 keV (Sguera et al., 2005). The emission processes
responsible for the very low stable emission from SFXTs and the
short  hysteric flares are probably explained by very wide
eccentric orbit (with a much longer period than classical SGXBs
${\sim}$15 days) and  in turn, with a low amount of wind material
available to accretion onto the compact companion. When the
collapsed star is closer to the supergiant donor star, highly
inhomogeneous, structured, wind clumps accretion can become
efficient and hance flares detectable by INTEGRAL in the soft
gamma ray range (Negueruela et al., 2005, In't Zand 2005). This
scenario implies a periodicity of the fast outbursts as they
should occur always relatively close to the periastron passage,
but to date periodicity has been observed only from one source:
IGR J11215$-$5952, with a period of ${\sim}$330 days (Sidoli et
al., 2006, Romano et al., 2007). In the light of these new and
exciting INTEGRAL results, the size of the population of SGXBs in
our Galaxy could have been severely underestimated. An entire
population of still undetected SFXTs could be hidden in our
Galaxy, on top the 8 firm source already discovered. Ongoing
observations with INTEGRAL are expected to provide new SFXTs
discovery and permit a breakthrough in our understanding of the
physical processes behind their fast X-ray transient behavior.
\begin{figure}[h]
 \centering
 \psbox[xsize=6.6cm]{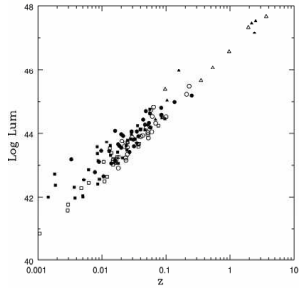}
\vspace{-4cm}
 \caption{Lum vs z for the
IBIS AGN sample.}
 \label{fig:AGN_2}
\end{figure}
\subsection{The extragalactic IBIS sky}
\begin{figure}[]
 \centering
 \psbox[xsize=6.5cm]{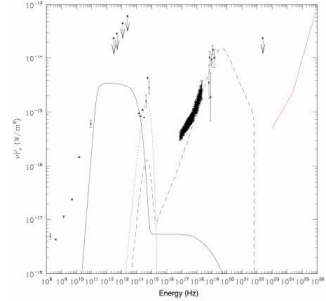}
\vspace{-4cm}
 \caption{SED of the QSO 4c04.42 (z=0.965)
suggestive of Bulk Comptonisation.}
 \label{fig:AGN_1}
\end{figure}

The INTEGRAL observing programme has been initially focused on the
Galaxy plane scanning and deep observation of the Galaxy center.
In recent years more attention has been dedicated to the
observation of the extragalactic sky, with particular attention to
obtain a substantial number of deep extragalactic field via the
so-called Key Programmes. The result is that from the third IBIS
survey (Bird et al. 2007) has provided a significant improvement
in the detection of extragalactic objects due to the larger sky
coverage available and deep observations ($>$2-5 Ms). One of the
main task was to have a better high energy picture of the close
Universe, i.e. the one amenable for a "complete sample"
observation with IBIS. In fact, quantifying the AGNs missing
fraction due to absorption (especially in the classical X-Ray band
Surveys) is necessary to fully understand the accretion evolution
of the Universe (Bassani et al., 2007a). This is also a
fundamental issue to understand the contribution of AGNs to the
X-ray cosmic diffuse background (Churazov et al., 2008) and in
order to test the current unified theories. This is only possible
with a good knowledge of the spectral shape and column density
distribution of the detected AGNs. Clearly, high energy
observations, above 10 keV, are necessary  to obtain an unbiased
measurement of the column densities and of the high energy cut-off
distributions of AGNs in the local Universe. Recently, this has
been possible thanks to the extragalactic INTEGRAL/IBIS deep
exposure and by the All Sky monitoring from SWIFT/BAT (Gehrels et
al., 2004) providing new complementary information on hard X-Ray
emitting AGNs. The list of cat3 AGNs has been obtained by the use
of a large number of identification and classification being
provided through optical spectroscopy and catalogue searches.
Within the sample of optically classified objects (Masetti et al.,
2006 a,b,c,d, 2008) about 120 are Seyfert galaxies and 13 are
blazars; within the Seyfert sample 58 objects are AGN of type
1-1.5 (circles in Figure \ref{fig:AGN_2}) while are 60 type 2
(squares), i.e. a ratio 1:1, which illustrates the power of
gamma-ray surveys to find narrow line AGN and 13 are blazars
(triangles), with about 10\% of the sample being radio loud
objects. The range of redshift probed by our sample is 0.001-3.668
while the 20-100 keV luminosities span from 10$^{41}$ to 10$^{48}$
erg $s^{-1}$. This, in turn, establish the limit sensitivity of
the cat3 Survey, i.e. about 1.5${\times}$10$^{11}$ erg cm$^{2}$
s$^{-1}$. This limit is improving with time in view of the largest
exposure for extragalactic fields and the basic insensitivity of
the IBIS observation to systematic error. The closest object
detected is the Seyfert 1.8 galaxy NGC4395 with a z=0.001 (Panessa
et al. 2006) and IGR J22517+2218 with a z=3.668 (Bassani et al.,
2007b). The IBIS sensitivity to faint extragalactic objects is
increasing with time providing, for the first time, the detection
of the brightest Blazars, the most powerful objects in the
observable Universe emitting from radio up to very high energy
gamma-ray. In the X/gamma-ray range, hard spectra are shown by the
highest luminosity objects (Fossati et al. 1998), and Flat
Spectrum Radio Quasars (FSRQ) are the most luminous class of
Blazars. The Spectral Energy Distribution (SED) of FSRQ exhibits
two main peaks (one between the IR and soft X-ray frequencies and
the other in the gamma-ray regime), disclosing the presence of two
main components: it is widely believed that the low energy one is
due to the Synchrotron radiation of relativistic electrons in a
jet, while the high energy one is due to Inverse Compton
scattering (IC) of the same electrons with a photon field
(Ghisellini et al. 1998). IBIS is becoming sensitive to bright
distant AGNs, and several of them are now becoming visible and
amenable to spectral studies (Table I). As an example, in Figure
\ref{fig:AGN_1} is shown the SED of the luminous QSO 4c04.42 at
z=0.965 for which has been observed and excess of emission below 2
keV (in the observer frame) suggesting Bulk Compton motion (De
Rosa et al.,
2008).\\
Conclusion

After more than 5.5 years of observations, INTEGRAL is still
performing a survey of the all sky and the deep monitoring of the
Galactic Plane, monitoring more than 420 sources and spending
several Ms observations in selected deep flieds; this is giving us
exciting results about new high energy emitters and discovering
new dozen of variable sources. The large field of view, the good
angular resolution and the deep observations provide all together
a powerful tool to discovery highly variable sources, like new
SFXTs, distant AGN flares, Blazar like IGR J22517+2218, the second
most distant one detected above 20 keV. The main conclusions are:
i) the number of INTEGRAL sources continue to increase
dramatically (+$>$ 200 sources in cat 4), ii) the HMXB numbers
increasing faster than LMXB after initial GCDE campaign, iii) AGN
numbers increased to ${\sim}$130 in cat3 and are almost doubled
today, boosted by exposure and follow-up campaigns, iv) the SFXT
have emerged as a new class: 14 (including candidates) have been
observed when not in outburst, v) a significant numbers of CVs are
now detected: almost exclusively IP sub-class, vi) PWN are now a
distinct class of soft gamma ray emitters. The Sky above 100 keV
is becoming more and more populated. Finally, a great expectation
is due to the launch of the GLAST high energy observatory just
placed in orbit the 11 June, 2008, during the MAXI Workshop. We
have clearly entered a "golden age" for high energy astrophysics
with contemporaneous in-flight operation of CHANDRA, XMM/Newton,
INTEGRAL, SWIFT, SUZAKU, AGILE, now GLAST and MAXI hopefully to
arrive soon.

Acknowledgement

The author wish a particular thanks to Prof. Masaru Matsuoka and
Prof. Nobuyuki Kawai for the kind invitation to participate to the
MAXI Workshop and for the invited talk on INTEGRAL. The author
acknowledge financial contribution from Contracts ASI I/008/07/0
and a special thanks to Mrs. Catia Spalletta for the professional
preparation of the manuscript.

\section{References}

Aharonian F., et al. 2005, Science 307, 1938  \\
Albert, J.,  et al., 2007, astro-ph0709.3763\\
Bassani, L., et al., 2007a astro-ph/0610455v1\\
Bassani, L. et al. 2007b Ap.J., 669,L1\\
Bazzano A., et al., 2006 ApJ, 649, L9\\
Bird, A.J., et al. 2004, ApJ, 607, 33\\
Bird, A.J., et al., 2006, ApJ, 636, 765\\
Bird, A.J., et al. 2007, ApJS, 170, 175\\
Bouchet, L. et al., 2008, ApJ, 679, 1315\\
Bodaghee A. et al., 2007, A\&A, 467,585\\
Churazov E. et al., 2008 A\&A, 467, 529\\
Courvoisier T., et al., 2003, A\&A, 411 L49\\
D'Alessio, V., et al., 2006, A\&A, 460, 653\\
De Rosa et al., MNRAS, 388, L54, 2008\\
Dean A. J., et al. 2008, MNRAS, submitted\\
Fossati, G., et al. 1998, MNRAS, 299, 433\\
Gehrels, N., et al. 2004, Ap.J., 611, 1005\\
Ghisellini, G., Celotti, A. et al., 1998, MNRAS, 301, 451\\
Lund N., et al. 2003, A\&A, 411, L231\\
Malizia A., et al. 2005, ApJL, 630, 157, \\
Masetti, N., et al. 2006a, A\&A, 448, 547 (Paper II)\\
Masetti, N., et al. 2006b, A\&A, 449, 1139 (Paper III)\\
Masetti, N., et al. 2006c, A\&A, 455, 11 (Paper IV)\\
Masetti, N.,  et al. 2006d, A\&A, 459, 21 (Paper V)\\
Masetti, N., et al., 2008, astro-ph/0802.0988 (Paper VI)\\
Mass-Hesse M. et al. 2003, A\&A, 411 L261\\
Negueruela, I. et al. 2005, ESA SP-604, 165\\
Panessa, F. et al. 2006 A.\&A, 455, 173\\
Romano et al., 2007, atel 1151 \\
Sguera, V., Barlow, E. J., et al., 2005, A\&A, 444, 221\\
Sguera, V., Bazzano, A., et al., 2006, ApJ, 646, 452\\
Sguera, V., Bazzano, A., et al., 2007, as-ph/0704.2737\\
Sguera, V., et al., 2008, as-ph/0805.0496\\
Sidoli, L., et al., 2006, A\&A, 450L, 9S\\
Skinner, G. \& P.Connell 2003, A\&A, 411, L123\\
Tavani et al., 2008, Cospar Symp.E12, in press\\
Ubertini P., et al. 2003, A\&A, 411, L131\\
Ubertini P., et al. 2005, ApJL, 29, 109\\
Ubertini P., De Rosa, A., Bazzano, A., et al., 2008, Nucl. Instr.\& Meth. A, 588, 63 \\
Vedrenne G,. et al. 2003 A\&A, 411, L63\\
Walter, R., et al. 2006, A\&A, 453, 133\\
Weidenspointner G. et al., Nature, 451,159,2008\\
Winkler C. et al. 2003, A\&A, 411, L1\\

\begin{table}[]
\scriptsize
 \caption{High-z INTEGRAL QSOs: the sample}
\begin{center}
\begin{tabular}{lccccc} \hline\hline\\[-6pt]
Source      &  z    &Reference   \\   \hline
 PKS1830-211   & 2.507     &    De Rosa et al. 2005, Zhang et al. 2008   \\
Swift J1656.3-3302    &  2.4&  Masetti et al. 2008       \\
  IGR J22517-2218   & 3.668   &   Bassani et al. 2007      \\
 QSO B0836-710    & 2.172 &   Bird et al. 2007  \\
   4C04.42    & 0.965   &    De Rosa et al. 2008     \\
   1RXS J19245   &   0.352&   Bird et al. 2007    \\
  IGR J03184-0014      & 1.93 &  Bird et al. 2007    \\
    QSO B0212+735     & 2.367   &  Bird et al. 2007      \\
  PKS 2149-307      & 2.345 &    Bird et al. 2007   \\
    PKS 0537-286      &  3.104&    Possibly detected in 20-200 keV     \\
 QSO 0723+679      & 0.884 &Possibly detected in 20-200 keV\\   \hline
\end{tabular}
\end{center}
\end{table}
\end{document}